\documentclass[sigconf,nonacm]{acmart}
\setcopyright{none}
\renewcommand\footnotetextcopyrightpermission[1]{}
\usepackage{graphicx}
\usepackage{booktabs}
\usepackage{microtype}
\usepackage{enumitem}
\setlist[enumerate]{topsep=2pt,itemsep=3pt,parsep=0pt,leftmargin=1.4em}
\newcommand{\claimhead}[1]{\par\addvspace{3pt}\noindent\textbf{#1}\hskip 0.5em\relax\ignorespaces}

\begin{document}

\title{The Assistant Erased You: Measuring Loss of Authorship Signals in
AI-Mediated Communication}

\author{Ushna Malik}
\affiliation{\institution{University of Iowa}\city{Iowa City}\state{Iowa}\country{USA}}
\email{mushna@uiowa.edu}
\author{Moiz Sadiq Awan}
\affiliation{\institution{Independent Researcher}\city{Mountain View}\state{California}\country{USA}}
\email{moizsawan@gmail.com}

\renewcommand{\shortauthors}{Malik and Awan}

\begin{abstract}
Research on AI-mediated communication has examined how AI assistance shapes
interpersonal perceptions and reduces stylistic diversity across users. We ask a
complementary question at the individual level: after a message is rewritten by an AI
writing assistant, can its author still be distinguished from others? We introduce
the \emph{Idiolect Erasure Rate} (IER), defined as the reduction in
authorship-attribution accuracy following AI-assisted rewriting. We evaluate IER on
three pre-generative-AI corpora using a stylometric model and the authorship-specific
LUAR model. Heavy rewriting substantially weakens authorship signals in personal
blogs and workplace email, reducing LUAR attribution by as much as 66.5 percentage
points, but has a much smaller effect on topic-structured news, where topic remains
predictive of authorship. Additional analyses suggest that rewriting produces
stylistic convergence despite substantial semantic overlap, and that
content-sensitive attributers understate the loss captured by authorship-specific
models. Heavily rewritten messages may also evade AI-text detectors, making them
difficult both to attribute to their human authors and to identify as AI-assisted, a
phenomenon we call \emph{double erasure}. IER measures computational attributability
rather than human recognition, and we release it as an open and reproducible protocol
for evaluating authorship-signal loss in AI-mediated communication.
\end{abstract}

\maketitle

\section{Introduction}

AI writing assistants increasingly mediate everyday written communication by
rewriting messages for clarity, professionalism, or tone before they are sent.
Research on AI-mediated communication (AIMC) has examined how such assistance affects
trust~\cite{jakesch2019,hohenstein2023}, responsibility~\cite{elish2019}, users'
beliefs~\cite{jakeschcowrite}, and stylistic diversity across
populations~\cite{padmakumar2024,agarwal2025}. We ask a complementary question at the
level of the individual: after a message passes through an AI writing assistant, can
its author still be distinguished from others?

Prior work in adversarial stylometry has shown that writers can deliberately modify
their language to evade computational attribution~\cite{brennan2012,potthast2016,krishna2023}.
We study a different setting. Rather than intentionally concealing identity, AI
writing assistants may weaken authorship signals as an unintended consequence of
rewriting text for clarity, fluency, or tone.

Writing carries recurring stylistic patterns, including word choice, sentence rhythm,
and punctuation, that together form an individual's \emph{idiolect}. These patterns
enable computational authorship-attribution systems to distinguish one writer from
another~\cite{koppel2009}. AI rewriting may alter these signals while preserving much
of the underlying message, raising the question of whether a person's writing remains
computationally attributable after assistance.

To study this phenomenon, we introduce the \emph{Idiolect Erasure Rate} (IER),
defined as the reduction in authorship-attribution accuracy following AI-assisted
rewriting. We evaluate IER across three communication registers using both
stylometric and neural attribution models. Our results show that AI rewriting
substantially weakens recoverable authorship signals in interpersonal writing, while
having much smaller effects where topical information remains highly predictive of
authorship.

Although we measure computational attribution rather than human recognition, we argue
that changes in attributable writing style represent an overlooked dimension of
AI-mediated communication. Understanding when AI assistance preserves or suppresses
these signals has implications for future identity-preserving writing assistants and
for how AI mediation is evaluated more broadly.

Our contributions are threefold:

\begin{enumerate}
\item We introduce the \emph{Idiolect Erasure Rate (IER)}, an open protocol for
quantifying the loss of computational authorship signals after AI-assisted rewriting.

\item We show that AI writing assistants substantially weaken authorship signals in
interpersonal communication while leaving topic-driven attribution largely intact in
topic-structured writing.

\item We identify \emph{double erasure}: AI-assisted text may simultaneously become
difficult to attribute to its human author and difficult to identify using current
AI-text detectors~\cite{sadasivan2023,krishna2023}.
\end{enumerate}

\section{Related Work}

\subsection{AI-mediated communication}

AI-mediated communication (AIMC) studies how AI systems modify, augment, or generate
messages on a person's behalf~\cite{hancock2020}. Prior work has shown that AI
mediation influences interpersonal trust~\cite{jakesch2019,hohenstein2023},
responsibility attribution~\cite{elish2019,hohenstein2020},
collaboration~\cite{hohenstein2023}, and the opinions users
express~\cite{jakeschcowrite}. More recently, researchers have examined how people
perceive AI-assisted writing and its implications for
authenticity~\cite{hwang2025,kadoma2025}. Our work examines a complementary question:
whether AI assistance changes the computationally attributable signals that
distinguish one writer from another.

\subsection{Authorship attribution and idiolect}

Computational authorship attribution exploits recurring stylistic patterns, including
function-word preferences, sentence rhythm, and punctuation, to identify writers
across anonymous texts~\cite{koppel2009,luar2021}. A related literature studies
authorship obfuscation, in which writers deliberately modify their style to evade
attribution~\cite{brennan2012,potthast2016}. Our setting differs in both intent and
mechanism. Rather than intentionally hiding identity, we study whether everyday
AI-assisted rewriting unintentionally weakens computationally measurable authorship
signals. We therefore adapt attribution accuracy as a way of quantifying stylistic
erosion rather than recovering authorship.

\subsection{From stylistic homogenization to individual identity}

Recent work has shown that AI writing assistants reduce stylistic diversity across
users~\cite{padmakumar2024,agarwal2025,doshi2024}, while recursive training similarly
reduces diversity within language models~\cite{shumailov2024}. These studies
characterize changes at the population level.

Communication research likewise argues that identity is expressed through language
and negotiated with an audience~\cite{goffman1959,ellison2012,hwang2025}. We
complement these perspectives by examining whether AI rewriting weakens the
computational signals through which an individual writer remains distinguishable.
Rather than measuring homogenization across a population, we quantify the loss of
attributable writing style for each author.

\section{Methodology}
\subsection{The Idiolect Erasure Rate}

The \emph{Idiolect Erasure Rate} (IER) quantifies how much AI-assisted rewriting
weakens computational authorship signals. We first train an authorship-attribution
model on each author's original writing, then evaluate whether it can still
identify the same authors after their messages have been rewritten by an AI
assistant. A larger reduction in attribution accuracy indicates that the assistant
has weakened the stylistic cues on which the attributer relies.

Formally, for attributer $f$, assistant $g_c$ under rewriting condition $c$, and
held-out human messages $\{x_i\}$ with corresponding authors $\{a_i\}$,
\[
\mathrm{IER}(g_c,f)=
\mathrm{acc}\!\big(f(\{x_i\}),\{a_i\}\big)
-
\mathrm{acc}\!\big(f(\{g_c(x_i)\}),\{a_i\}\big),
\]
where IER is the percentage-point reduction in attribution accuracy after
AI-assisted rewriting.

We report results using both a \textbf{surface} (stylometric) attributer and a
\textbf{deep} (neural) attributer, treating agreement and disagreement between them
as informative. Because attribution performance depends on both the evaluation
setting and the attribution method, IER is not an intrinsic property of an AI
assistant but of the assistant, rewriting condition, attributer, and corpus
considered together.

\subsection{Experimental setup}

\textbf{Corpora.}
We evaluate IER on three corpora representing different communication registers.
The \emph{Blog Authorship Corpus}~\cite{schler2006} contains informal, personal,
topic-diverse writing (50 authors, 200 held-out messages), making it representative
of the register targeted by AI writing assistants and one in which topic is only
weakly associated with author identity. The \emph{Enron Email Corpus} consists of
authentic interpersonal email (20 users, 80 held-out messages), representing a more
formulaic communication register. The \emph{Reuters C50} corpus contains news
articles in which each journalist writes within a fixed topical beat (25 authors,
100 held-out messages), providing a control condition where topic strongly
correlates with authorship. All three corpora predate modern generative AI systems.

Rather than truncating documents, we retain messages at their natural length (mean
190 words, capped at 400). Truncation artificially reduces attribution performance
and produces incomplete fragments that AI assistants partially reconstruct,
introducing an additional confound.

\textbf{Attributers.}
We evaluate both a \emph{surface} and a \emph{deep} authorship attributer. The
surface model uses TF--IDF character ($2$--$4$)-grams and word ($1$--$2$)-grams with
a linear SVM. The deep model is LUAR~\cite{luar2021}, which represents each author
by the mean embedding of their training documents and assigns authorship using
nearest-profile retrieval.

A style-sensitive attributer is essential for measuring idiolect erosion. MiniLM is
largely insensitive to word-order perturbations (accuracy drops only from $0.385$ to
$0.360$ after random word shuffling), indicating that it relies primarily on
semantic content. In contrast, LUAR drops from $0.710$ to $0.205$ under the same
manipulation, demonstrating substantially greater sensitivity to stylistic
information. We therefore use LUAR as our primary deep attributer and report MiniLM
only as a topic-sensitive baseline.

\textbf{Assistant.}
Our primary rewriting model is the local open-weight model Qwen2.5-1.5B-Instruct,
decoded greedily to ensure deterministic and reproducible outputs. Our evaluation
pipeline is model-agnostic and also supports commercial AI assistants
(\S\ref{sec:results}).

\textbf{Rewriting conditions.}
We evaluate three prompting conditions: \emph{light}, which corrects grammar and
spelling only; \emph{heavy}, which rewrites text for clarity and professionalism;
and \emph{preserve}, which improves writing while explicitly preserving the
author's voice. All prompts are released with the code.

\textbf{Evaluation protocol.}
For each author, a disjoint set of messages is held out for evaluation, while the
remaining documents are used for training, providing the attributer with as much
data as each corpus permits (blogs: mean 119 posts per author; LUAR profiles average
60 training documents). IER compares attribution accuracy on the held-out original
messages with accuracy on their AI-rewritten counterparts.

Authorship attribution is evaluated in the closed-set setting, where every test
message belongs to one of the known authors. This provides an upper bound on
attribution performance, whereas real-world recognition is generally open-set.
Chance performance is therefore the reciprocal of the number of authors.

Authors are selected according to document availability. Random author subsampling
yields comparable attribution baselines ($0.80$--$0.88$ on the blog corpus),
indicating that this selection strategy does not materially inflate performance.

Statistical significance is assessed using McNemar's exact test. We report 95\%
bootstrap confidence intervals and, for headline results, cluster-bootstrap
confidence intervals over authors. To facilitate reproducibility, we release all
prompts, random seeds, LUAR checkpoints, and author splits.\footnote{Code and data:
the IER protocol, prompts, seeds, the LUAR checkpoint, and the per-corpus author
splits are available at \url{https://github.com/ushnamalikk/idiolect-erasure-rate}.}

\begin{figure*}[t]
\centering
\includegraphics[width=0.86\textwidth]{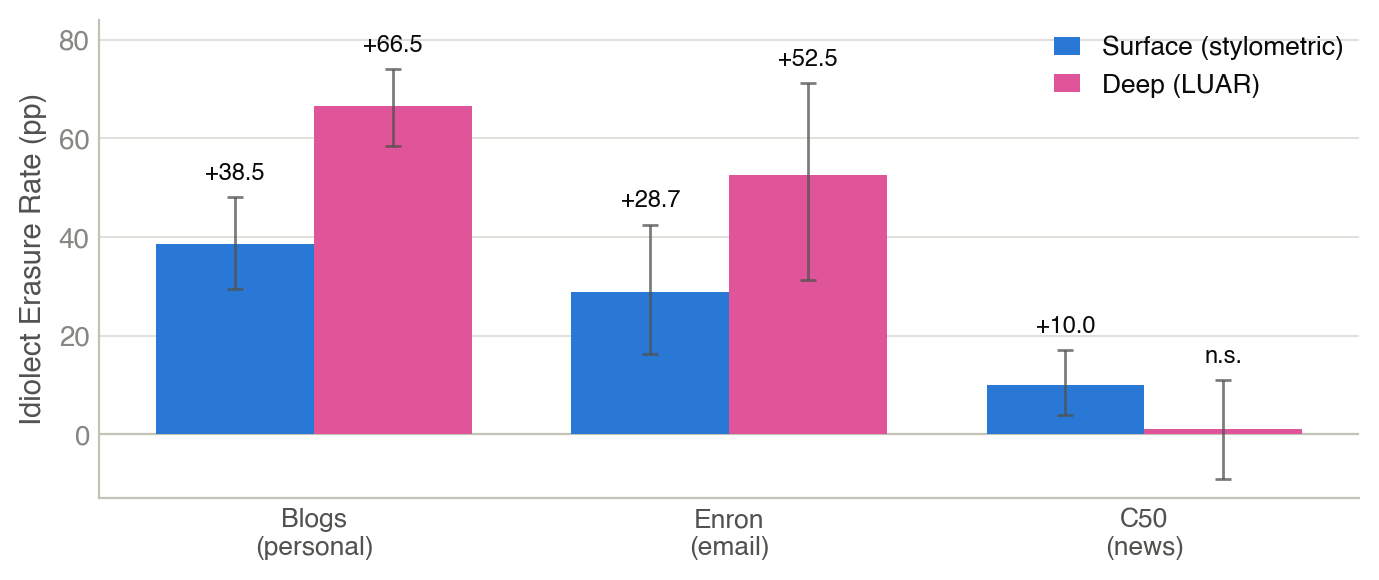}
\caption{Heavy-rewrite IER across three corpora (matching Table~\ref{tab:results}),
for the surface (stylometric) and deep (LUAR) attributers, with author-clustered
95\% CIs. Surface attribution is significantly erased on every register; the deep
fingerprint is erased strongly on both interpersonal corpora (Blogs $+66.5$, Enron
$+52.5$) but is null on beat-structured news (C50), where topic stands in for
identity. Bars are significant ($p<0.001$) unless marked n.s.}
\label{fig:ier}
\end{figure*}

\section{Results}\label{sec:results}

\begin{table}[t]
\centering
\caption{Heavy-rewrite IER in percentage points, with the baseline attribution
accuracy it is measured from (base). Surface is stylometric; deep is LUAR.
Post-rewrite accuracy is base${}-{}$IER${}/100$. $\ast$: $p<0.001$; n.s.: not
significant. $n=200$ (blogs), $80$ (Enron), $100$ (C50).}
\label{tab:results}
\setlength{\tabcolsep}{4pt}
\begin{tabular}{lccccc}
\toprule
 & & \multicolumn{2}{c}{surface} & \multicolumn{2}{c}{deep (LUAR)} \\
\cmidrule(lr){3-4}\cmidrule(lr){5-6}
Corpus & chance & base & IER & base & IER \\
\midrule
Blogs (personal) & .02 & .810 & $+38.5^{\ast}$ & .815 & $+66.5^{\ast}$ \\
Enron (email)    & .05 & .938 & $+28.7^{\ast}$ & .713 & $+52.5^{\ast}$ \\
C50 (news)       & .04 & .910 & $+10.0^{\ast}$ & .710 & $+1.0$ (n.s.) \\
\bottomrule
\end{tabular}
\end{table}

\begin{figure}[t]
\centering
\includegraphics[width=\linewidth]{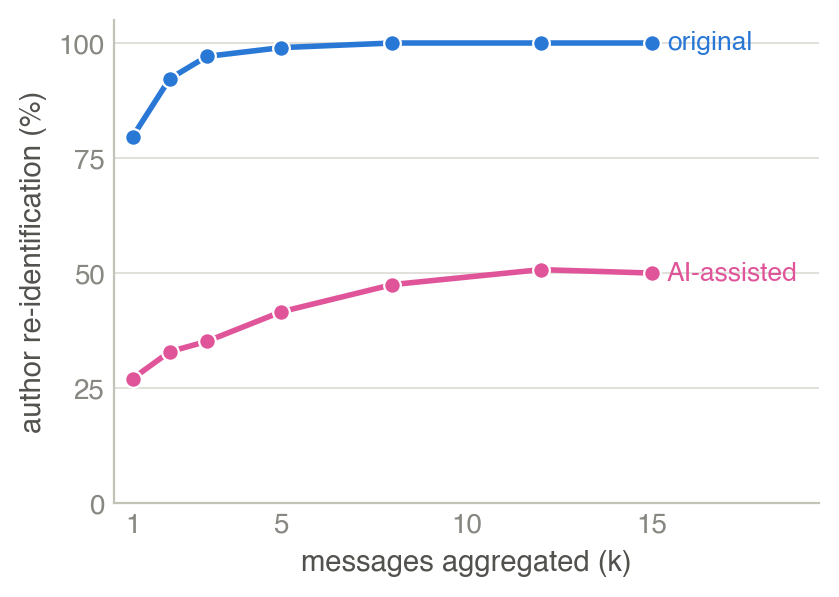}
\caption{Population accumulation (heavy rewrite, LUAR, 20 blog authors).
Re-identification from pooled messages reaches $100\%$ by $k{=}8$ for originals but
saturates near $50\%$ for assisted messages, so aggregation does not recover erased
identity.}
\label{fig:accumulation}
\end{figure}

\claimhead{Strong authorship signals before AI assistance.}
Before rewriting, both surface and deep attributers recover authors far above
chance across all three corpora (Table~\ref{tab:results}). Surface attribution
reaches 0.810, 0.938, and 0.910 on Blogs, Enron, and Reuters C50, respectively,
while LUAR achieves 0.815, 0.713, and 0.710. In the closed-set setting, aggregating
multiple original messages identifies the correct author with high reliability
(Fig.~\ref{fig:accumulation}). The central question is how much of this authorship
signal survives AI-assisted rewriting.

\claimhead{Surface authorship signals weaken consistently.}
Heavy AI rewriting significantly reduces surface attribution across all three
corpora (Fig.~\ref{fig:ier}; Table~\ref{tab:results}). The largest reduction occurs
on the Blog corpus (IER = +38.5 points), followed by Enron (+28.7) and Reuters C50
(+10.0). Grammar-only rewriting produces substantially smaller effects, suggesting
that extensive stylistic rewriting, rather than simple error correction, is
primarily responsible for erasing surface authorship cues.

\claimhead{Deep authorship signals are substantially more vulnerable.}
The effect is even stronger for deep attribution. Under heavy rewriting, LUAR loses
66.5 percentage points on Blogs and 52.5 on Enron, corresponding to the loss of
more than three quarters of the recoverable authorship signal. The effect grows
monotonically with rewriting intensity, and even prompts that explicitly instruct
the assistant to preserve the author's voice remove most of the recoverable signal.
These findings remain significant under author-level cluster bootstrapping.

\claimhead{The observed loss reflects stylistic convergence.}
Several analyses indicate that the measured erasure reflects changes in writing
style rather than changes in semantic content. First, rewritten texts become
substantially less distinguishable from one another, indicating convergence toward
more similar stylistic representations. Second, semantic similarity between original
and rewritten messages remains high despite large reductions in attribution
accuracy. Third, function-word attribution, which minimizes lexical-content
information, also exhibits substantial degradation. Together, these results suggest
that AI rewriting primarily weakens stylistic rather than semantic cues.

\claimhead{The choice of attributer matters.}
IER depends strongly on the attribution model. Although the stylometric model and
LUAR achieve nearly identical baseline accuracy, LUAR exhibits substantially larger
erasure after rewriting, indicating that IER depends on the type of authorship
signal an attributer captures rather than its overall accuracy. In contrast,
content-sensitive models such as MiniLM consistently underestimate erasure because
they rely more heavily on semantic information than on stylistic patterns.

The Reuters C50 corpus illustrates this distinction. Although LUAR achieves strong
baseline attribution performance, heavy rewriting produces little deep erasure
because topic remains highly predictive of authorship. Function-word attribution
nevertheless declines substantially, indicating that stylistic signals are weakened
even when topical information continues to support attribution.

\claimhead{The effect generalizes across AI assistants.}
Comparable levels of deep authorship erasure are observed for commercial assistants,
including GPT-4o-mini and Gemini Flash, as well as across multiple sizes of the
Qwen2.5 family. The phenomenon therefore does not appear to depend on a particular
model architecture or parameter scale.

\claimhead{Identity loss accumulates across correspondence.}
Aggregating more rewritten messages does not recover a writer's identity. Whereas
attribution from original messages approaches perfect accuracy after observing
several messages, attribution from rewritten messages saturates at approximately
50\% and shows little improvement thereafter (Fig.~\ref{fig:accumulation}). This
suggests that idiolect erasure is best understood as an accumulation across a
person's correspondence rather than repeated degradation of an individual message.

\claimhead{Voice-preservation prompts are insufficient.}
Explicitly instructing the assistant to preserve the author's voice reduces
surface-level erasure but leaves most of the deep authorship signal unrecovered.
Simple prompting therefore appears insufficient to preserve computationally
identifiable writing style, motivating future work on identity-preserving AI writing
assistance.

\claimhead{Robustness checks.}
All reported effects survive a Bonferroni correction across our tests (only C50
light does not, $p=0.031$), and every significant IER excludes zero under
author-clustered bootstrapping (deep intervals $[+58,+74]$ on Blogs, $[+31,+71]$ on
Enron). Grammar-only (light) rewriting erases far less than heavy rewriting (surface
$+5.0$ on Blogs; $+3.7$, n.s., on Enron), and a non-generative grammar-correction
baseline removes only $+15$ LUAR points versus $+66.5$ for generative rewriting, so
the effect is specific to generative rewriting rather than any edit. Enron's high
baseline is partly signature leakage: stripping sign-off names and quoted text
lowers baselines (surface $0.94$ to $0.85$, LUAR $0.71$ to $0.58$) but the erasure
persists ($+30$/$+39$, $p<10^{-6}$), while the same stripping leaves Blog baselines
unchanged. Exact commercial IERs (heavy, LUAR) are GPT-4o-mini $+58.5$/$+53.8$ and
Gemini $+60.5$/$+52.5$ on Blogs/Enron.

\section{Discussion}

Our findings suggest that computational authorship signals constitute an overlooked
dimension of AI-mediated communication, with implications for the design and
evaluation of AI writing assistants.

\claimhead{Design implications.}
IER provides both a design objective and an evaluation metric for
identity-preserving AI writing assistance. Rather than relying on prompting alone,
future systems could explicitly optimize for preserving computational authorship
signals while improving writing quality. Our \emph{preserve} condition suggests that
prompting alone is insufficient, particularly for deeper stylistic representations.
More generally, IER could serve as a measurable ``fingerprint budget'' that allows
users to control how much of their writing style is retained.

Disclosure alone is not enough. Labels such as ``AI-assisted'' indicate provenance,
not whether a message still reflects its author's distinctive style. Moreover,
heavily rewritten messages may evade current AI-text
detectors~\cite{sadasivan2023,krishna2023}, producing what we term \emph{double
erasure}: text that is difficult both to attribute to its human author and to
identify as AI-assisted.

\claimhead{Implications for AIMC.}
Existing AIMC research has examined how AI mediation affects trust, responsibility,
beliefs, and stylistic diversity. Our results suggest a complementary perspective: AI
assistance may also weaken the attributable stylistic signals that distinguish one
writer from another. We therefore propose idiolect preservation as an additional
dimension of AI-mediated communication. Its desirability, however, is
context-dependent: preserving a recognizable voice may support authenticity and
accountability, while reduced attributability may benefit privacy or anonymity.
Future work should therefore examine how systems can balance writing quality, user
control, and authorship preservation across different communication settings.

\section{Limitations and Conclusion}

IER measures computational attributability, not human recognition. Reduced
attribution therefore indicates weaker authorship signals, not necessarily lower
recognition by familiar readers. Future work should include human-recognition studies
and edit-matched human baselines.

We evaluate single-pass rewriting by three assistants across three registers, with
20--50 authors per corpus. Broader, multilingual, and longitudinal evaluations are
needed. Erasure may also benefit privacy-sensitive settings, although the effect
remains consistent across assistants, attributers, and rewriting conditions.

Taken together, our findings show that AI-assisted rewriting weakens authorship
signals in interpersonal communication, promotes stylistic convergence across
writers, and leaves topic-driven attribution largely intact. We introduce the
Idiolect Erasure Rate as a reproducible framework for quantifying this phenomenon and
supporting future research on writing assistants that better preserve individual
style.


\newpage
\begin{thebibliography}{99}
\bibitem{agarwal2025} D.~Agarwal, M.~Naaman, and A.~Vashistha. 2025. AI
Suggestions Homogenize Writing Toward Western Styles and Diminish Cultural
Nuances. In \emph{CHI}.
\bibitem{brennan2012} M.~Brennan, S.~Afroz, and R.~Greenstadt. 2012. Adversarial
Stylometry: Circumventing Authorship Recognition to Preserve Privacy and
Anonymity. \emph{ACM TISSEC} 15(3).
\bibitem{doshi2024} A.~R.~Doshi and O.~P.~Hauser. 2024. Generative AI enhances
individual creativity but reduces the collective diversity of novel content.
\emph{Science Advances} 10(28).
\bibitem{potthast2016} M.~Potthast, M.~Hagen, and B.~Stein. 2016. Author
Obfuscation: Attacking the State of the Art in Authorship Verification. In
\emph{CLEF (Working Notes)}, PAN.
\bibitem{elish2019} M.~C.~Elish. 2019. Moral Crumple Zones: Cautionary Tales in
Human-Robot Interaction. \emph{Engaging Science, Technology, and Society} 5,
40--60.
\bibitem{ellison2012} N.~B.~Ellison, J.~T.~Hancock, and C.~L.~Toma. 2012. Profile
as Promise: A Framework for Conceptualizing Veracity in Online Dating
Self-Presentations. \emph{New Media \& Society} 14(1).
\bibitem{goffman1959} E.~Goffman. 1959. \emph{The Presentation of Self in
Everyday Life}. Doubleday.
\bibitem{hancock2020} J.~T.~Hancock, M.~Naaman, and K.~Levy. 2020. AI-Mediated
Communication: Definition, Research Agenda, and Ethical Considerations.
\emph{JCMC} 25(1).
\bibitem{hohenstein2020} J.~Hohenstein and M.~Jung. 2020. AI as a Moral Crumple
Zone: The Effects of AI-Mediated Communication on Attribution and Trust.
\emph{Computers in Human Behavior} 106.
\bibitem{hohenstein2023} J.~Hohenstein et al. 2023. Artificial intelligence in
communication impacts language and social relationships. \emph{Scientific
Reports} 13:5487.
\bibitem{hwang2025} A.~H.-C.~Hwang, Q.~V.~Liao, S.~L.~Blodgett, A.~Olteanu, and
A.~Trischler. 2025. ``It was 80\% me, 20\% AI'': Seeking Authenticity in
Co-Writing with Large Language Models. \emph{Proc. ACM Hum.-Comput. Interact.}
(CSCW).
\bibitem{jakesch2019} M.~Jakesch, M.~French, X.~Ma, J.~T.~Hancock, and
M.~Naaman. 2019. AI-Mediated Communication: How the Perception that Profile
Text was Written by AI Affects Trustworthiness. In \emph{CHI}.
\bibitem{jakeschcowrite} M.~Jakesch, A.~Bhat, D.~Buschek, L.~Zalmanson, and
M.~Naaman. 2023. Co-Writing with Opinionated Language Models Affects Users'
Views. In \emph{CHI}.
\bibitem{kadoma2025} K.~Kadoma, D.~Metaxa, and M.~Naaman. 2025. Generative AI and
Perceptual Harms: Who's Suspected of Using LLMs? In \emph{CHI}.
\bibitem{koppel2009} M.~Koppel, J.~Schler, and S.~Argamon. 2009. Computational
methods in authorship attribution. \emph{JASIST} 60(1).
\bibitem{krishna2023} K.~Krishna, Y.~Song, M.~Karpinska, J.~Wieting, and
M.~Iyyer. 2023. Paraphrasing Evades Detectors of AI-Generated Text, but Retrieval
is an Effective Defense. In \emph{NeurIPS}.
\bibitem{padmakumar2024} V.~Padmakumar and H.~He. 2024. Does Writing with
Language Models Reduce Content Diversity? In \emph{ICLR}.
\bibitem{luar2021} R.~Rivera-Soto et al. 2021. Learning Universal Authorship
Representations. In \emph{EMNLP}.
\bibitem{sadasivan2023} V.~S.~Sadasivan, A.~Kumar, S.~Balasubramanian, W.~Wang,
and S.~Feizi. 2023. Can AI-Generated Text be Reliably Detected? \emph{arXiv preprint
arXiv:2303.11156}.
\bibitem{schler2006} J.~Schler, M.~Koppel, S.~Argamon, and J.~Pennebaker. 2006.
Effects of Age and Gender on Blogging. In \emph{AAAI Spring Symposium}.
\bibitem{shumailov2024} I.~Shumailov et al. 2024. AI models collapse when
trained on recursively generated data. \emph{Nature} 631.
\end{thebibliography}
\end{document}